\documentclass[conference]{IEEEtran}
\usepackage{cite}
\usepackage{amsmath,amssymb,amsfonts}
\usepackage{algorithmic}
\usepackage{graphicx}
\usepackage{textcomp}
\usepackage[table]{xcolor}
\usepackage{float}
\usepackage{caption}
\usepackage{subcaption}
\usepackage{empheq}
\usepackage{colortbl}
\usepackage{color,soul}
\usepackage{hhline}
\usepackage{siunitx}
\usepackage{booktabs}
\usepackage{url}
\usepackage[none]{hyphenat}

\usepackage[a4paper, total={184mm,239mm}]{geometry}
\def\BibTeX{{\rm B\kern-.05em{\sc i\kern-.025em b}\kern-.08em
    T\kern-.1667em\lower.7ex\hbox{E}\kern-.125emX}}

\usepackage{multirow}

\usepackage{float}

\begin{document}

%-----------------------------------------------------------
% Paper title
% ----------------------------------------------------------

%\title{On the Automatic Generation of GPU-like Accelerators for ASICs
%}
% M: Que tal
%% TITULO ORIGINAL
\title{G-GPU: A Fully-Automated Generator of GPU-like ASIC Accelerators\\[-1.1ex]}

%% FOR THE FUTURE!!!
%\title{A Framework for Generating Domain-specific ASIC Accelerators based on GPU Architectures\\[-1.1ex]}

% AGPU: A Fully-Automated Generator of GPU-like Accelerators for ASIC Flows

\author{
  \IEEEauthorblockN{
    Tiago D. Perez\IEEEauthorrefmark{1},
    Márcio M. Gonçalves\IEEEauthorrefmark{2},
    Leonardo Gobatto\IEEEauthorrefmark{2},
    Marcelo Brandalero\IEEEauthorrefmark{3},
    José Rodrigo Azambuja\IEEEauthorrefmark{2},\\
    Samuel Pagliarini\IEEEauthorrefmark{1}
  }
  
  \IEEEauthorblockA{\IEEEauthorrefmark{1}
    Department of Computer Systems, Tallinn University of Technology (TalTech), Estonia
  }
  
  \IEEEauthorblockA{\IEEEauthorrefmark{2}
    Institute of Informatics, Federal University of Rio Grande do Sul (UFRGS), Brazil
  }
  
  \IEEEauthorblockA{\IEEEauthorrefmark{3}
    Brandenburg University of Technology (B-TU), Germany
  }
  \small Emails:\{tiago.perez,samuel.pagliarini\}@taltech.ee,\{marcio.goncalves,leonardo.gobato,jose.azambuja\}@inf.ufrgs.br,marcelo.brandalero@b-tu.de
  \\[-30pt]
}

%{\small Emails: \{tiago.perez,samuel.paglianiri\}@taltech.ee,	\{marcio.goncalves,leonardo.gobatto,jose.azambuja\}@inf.ufrgs.br, marcelo.brandalero@gmail.com}\\[-15pt]

\maketitle

% ----------------------------------------------------------% Abstract of the paper
%-----------------------------------------------------------
\begin{abstract}
Modern Systems on Chip (SoC), almost as a rule, require accelerators for achieving energy efficiency and high performance for specific tasks that are not necessarily well suited for execution in standard processing units. Considering the broad range of applications and necessity for specialization, the design of SoCs has thus become expressively more challenging. In this paper, we put forward the concept of G-GPU, a general-purpose GPU-like accelerator that is not application-specific but still gives benefits in energy efficiency and throughput. Furthermore, we have identified an existing gap for these accelerators in ASIC, for which no known automated generation platform/tool exists. Our solution, called GPUPlanner, is an open-source generator of accelerators, from RTL to GDSII, that addresses this gap. Our analysis results show that our automatically generated G-GPU designs are remarkably efficient when compared against the popular CPU architecture RISC-V, presenting speed-ups of up to 223 times in raw performance and up to 11 times when the metric is performance derated by area. These results are achieved by executing a design space exploration of the GPU-like accelerators, where the memory hierarchy is broken in a smart fashion and the logic is pipelined on demand. Finally, tapeout-ready layouts of the G-GPU in 65nm CMOS are presented.

\end{abstract}
% ----------------------------------------------------------% Index terms
%-----------------------------------------------------------

\section{Introduction} \label{sec:introduction}

New computer applications, especially in the field of Artificial Intelligence (AI), keep pushing the need for more energy-efficient hardware architectures~\cite{ieee_tutorial}. For many years, application- and domain-specific accelerators, designed by specializing to the task at hand, have been the standard choice for achieving high energy efficiency. Canonical examples are crypto cores for efficient encryption/decryption \cite{aes_date} and Graphics Processing Unit (GPUs) for which even specialized programming languages and paradigms have been proposed \cite{hicuda}.

GPU architectures focus on specialized massively parallel many-core processors that take advantage of Thread-Level Parallelism (TLP) to handle highly parallelizable applications in a Single-Instruction Multiple Threads (SIMT) paradigm. GPUs, as the name implies, have been traditionally designed for graphics applications but have recently evolved into efficient general-purpose accelerators for High-Performance Computing (HPC). HPC applications have a wide range, including oil exploration, bioinformatics, and the thriving AI and Machine Learning (ML) domains~\cite{deepdeep}. NVIDIA GPUs, for instance, are used as accelerators in several top500 supercomputers.

However, despite its widespread use as accelerators, research in GPU architectures is limited due to the lack of open-source models at a sufficiently low level of abstraction and that are representative of modern architectures. To the best of our knowledge, the only configurable open-source GPU architectures available in the literature are FlexGripPlus~\cite{FlexGripPlus} and FGPU~\cite{fpgu_paper}. The first is based on the NVIDIA G80 decade-old architecture and has never been deployed to an FPGA board. The second was designed specifically for FGPA platforms. Therefore, the literature has not yet tackled the challenges in designing, configuring, and implementing modern GPU architectures for ASICs -- a platform that presents challenges that are far from those in FPGA design. Still, all commercial GPUs are designed as ASICs.

This work proposes to \textbf{bridge this gap} with GPUPlanner, an automated and open-source framework for generating ASIC-specific GPU-like accelerators as IP. We term these general-purpose accelerators G-GPUs. GPUPlanner helps designers in generating GPU-like accelerators through user-driven customization and automated physical implementation. Customization is performed according to a given GPU architecture through a series of parameters that define computation characteristics (e.g., number of processing units) and memory access (e.g., cache sizes), thus providing designers a high degree of scalability to better fit the generated IP into their systems. Implementation strategies explore the use of \emph{smart memories} and on-demand pipeline insertion.

We evaluate our proposed framework by implementing four flavors of G-GPU architectures in terms of performance, power, and area (PPA). Additionally, we provide a reasonable comparison with the popular CPU architecture RISC-V~\cite{riscv_ref, riscv-github} in terms of raw performance speed-up and performance speed-up derated by area. The findings from our experiments can help designers in better understanding how G-GPU can be used as an accelerator in their system, as well as the expected performance gains. In summary, our main contributions are:

\begin{itemize}
% Comentario Sam: one of the claims should be the open source nature of our tool
    \item GPUPlanner, an open-source framework for automated generation of GPU-like accelerators, from RTL to GDSII.
    
    \item G-GPU, a domain-specific ASIC accelerator based on general-purpose GPU-like architectures.
    
    \item Design space exploration for performance, power, and area evaluations of G-GPUs generated by GPUPlanner.
    
    \item Rich results and trade-offs encountered during logical and physical synthesis in a commercial 65nm CMOS technology.% commercial technology.
    
\end{itemize}

\section{Hardware Accelerators and our Baseline GPU}

This section discusses hardware accelerators and the baseline GPU architecture chosen for generating multiple versions of G-GPU. We tackle three classes of hardware accelerators and discuss in detail FGPU and its customization capabilities.

\subsection{Hardware Accelerators}

Domain-specific hardware accelerators can provide orders of magnitude speed-up and energy efficiency over general-purpose processing architectures. However, they must be manually tailored to most efficiently tackle domain-specific characteristics and extract the most efficiency from a given application domain. 

Recent developments in High-Level Synthesis (HLS) are encouraging and have helped in bridging this gap for many -- but not all -- application domains. The fact is that HLS ~\cite{HLS} is a fantastic early prototyping approach that trades off some performance for flexibility. Yet, for ASIC designs, this trade-off is not interesting, or the performance is insufficient~\cite{DSAGEN}. It remains that accelerators that target ASIC platforms are optimized to a tee, a process that is costly and time-consuming.

In a nutshell, domain- or application-specific accelerators cost too much; the design and implementation of their hardware and software-stack cannot be easily justified. This scenario presents itself as an opportunity where general-purpose accelerators have gained ground. These accelerators benefit from modern programming languages that have effective supporting tools for programming, debugging, and deployment. 

%% OUR APPROACH
Our proposed GPUPlanner framework combines the efficiency from domain-specific accelerators and the ease of use (i.e., programmability) from general-purpose architectures into G-GPU. The result is an automatically generated domain-specific ASIC accelerator based on GPU architectures that can be easily programmed with modern programming languages. This is the main contribution of our work.

\subsection{FGPU: our Baseline GPU Architecture} \label{sec:arch}

FGPU is a configurable open-source GPU-like soft processor designed to accelerate workloads that fit in the SIMT paradigm~\cite{fpgu_paper}. It was originally conceptualized and designed for FGPA platforms. However, its RTL design description can be ported to ASIC platforms with precise adaptations to its memory hierarchy. This GPU-like architecture has a supporting LLVM-based compiler, which can compile existing OpenCL kernels and provides designers with the ability for fast software development, debugging, and deployment. Most importantly, FGPU can be artlessly scaled up to 64 processing units (and beyond with additional support) and is deeply configurable in terms of operations, instructions, and memory access.

\begin{figure}[tb]
    \centering
    \includegraphics[width=\linewidth]{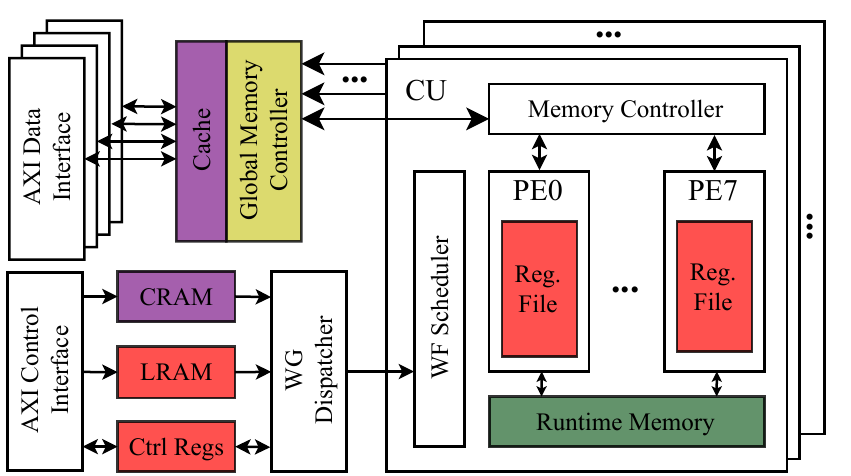}
    \caption{FGPU architecture with memories colored according to the layouts displayed in Figs. 3 and 4.}
    \label{fig:fgpu_arch}
\end{figure}

Fig.~\ref{fig:fgpu_arch} presents an overview of FGPU's architecture. Its main component is the Compute Unit (CU), a SIMD machine of 8 identical Processing Elements (PE0 - PE7) that can be spatially replicated up to eight times. A single CU can run up to 512 work-items (a computational kernel in OpenCL) and supports full thread-divergence, i.e., each work-item is allowed to take a different path in the control flow graph. Work-items are grouped into Wavefronts (WFs) that execute concurrently in a CU, and WFs are combined into Workgroups (WGs), which share a program counter and are assigned to a CU. FGPU is also deeply pipelined. The size of these parameters is entirely configurable when implementing the design.

FGPU features a Runtime Memory (RTM) and a Data cache. The cache is a central, direct-mapped, multi-port, and write-back system that can serve multiple read/write requests simultaneously. Besides, it integrates numerous data movers that can parallelize the data traffic on up to four AXI Data interfaces. The whole architecture is controlled on the hardware side through a single AXI Control interface. On the software side, only standard OpenCL-API procedures are needed. These parameters can also be configured in size and number.

Several past works have modified the FGPU to adapt it to different application domains. In \cite{fgpu_deep_learning}, the authors have included new instructions along with micro-architecture and compiler enhancements to specialize FPGU for persistent deep learning, achieving  56–693x speed-up in PDL applications. However, the modifications are not made publicly available.

MIAOW \cite{miow_gpgpu} is GPU-like implementation based on the AMD Southern Islands architecture and supporting its ISA. However, it is not fully synthesizable since the register files, on-chip networks, and memory controllers are described using behavioral C/C++. Scratch \cite{scratch_gpgpu} extended MIAOW with automatic identification of the specific requirements of each application kernel and a tool that allows for the generation of application-specific and FPGA-implementable trimmed-down GPU-inspired architectures. However, it targets FPGAs rather than ASICs, and its source code is not publicly available, limiting its use to the community.

To the best of our knowledge, this is the first work in the literature to propose a tool that automatically generates tapeout-ready domain-specific accelerators based on GPU-like architectures and makes it publicly available. Moreover, our framework enables a novel and comprehensive design-space exploration of the proposed design w.r.t. logic and memory components, which must be finely adjusted for the best efficiency.
Compared to related works on the FGPU, we target ASIC flows and conduct a DSE of different parameters for the memory hierarchy, significantly increasing the design complexity over FPGA design.
Compared to MIAOW and SCRATCH, our design and framework are fully synthesizable and tapeout-ready (RTL-to-GDSII) and available to the community for further investigations.

\begin{figure}[t]
    \centering
    \includegraphics[width=0.75\linewidth]{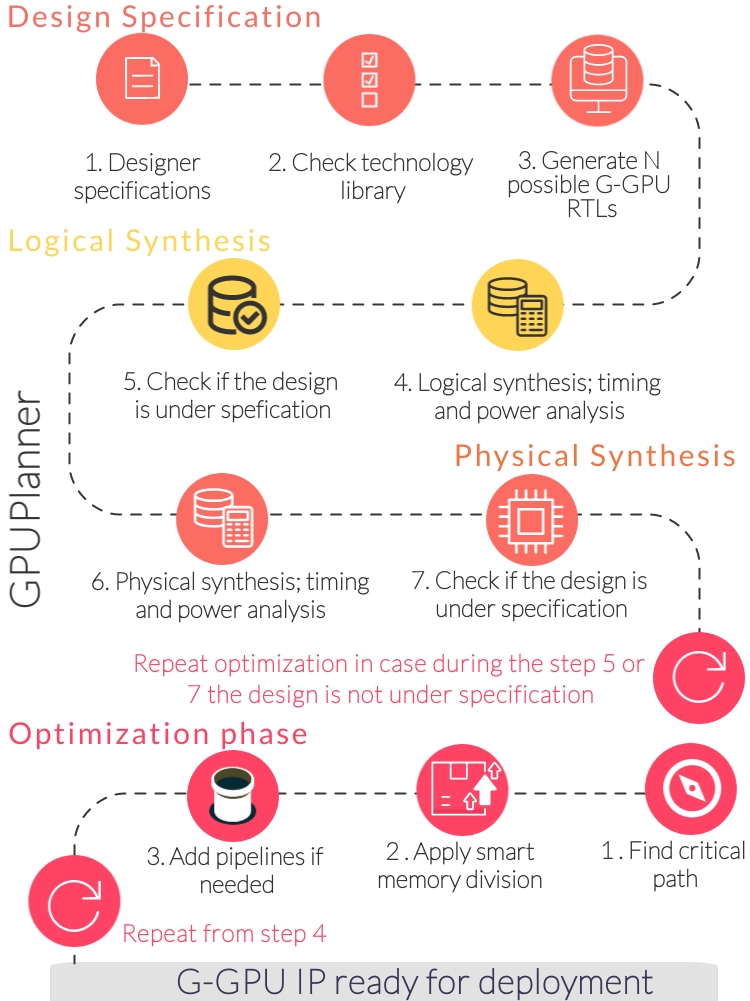}
    \caption{GPUPlanner's G-GPU generation flow.}
    \label{fig:ip_flow}
\end{figure}
    
\section{GPUPlanner Framework}

Our experimental investigation started from migrating the FGPU, originally designed for FGPA, to ASIC. To this end, a few changes in the architecture were necessary. As compilers for FGPA have a feature to infer memory from RTL automatically, all the memory blocks in the FGPU code were described as regular FFs. In ASIC, memory IPs are hand-instantiated instead of inferred. Thus, the first task was to clearly define intended behavior from the code and instantiate memory modules. In our experiments, we utilized a 65nm commercial technology. Its memory compiler offers single and dual-port low-power SRAM, with parameters ranging from 16-65536 for addresses and 2-144 bits for word size.

 One of our main goals is to achieve the best performance, power, and area (PPA) ratio possible from the G-GPU, exercising the maximum possible design space. The result of this was a selection of versions for different scenarios. The first aspect analyzed was the performance. This is done by finding the maximum operating frequency, which does not violate timing. For the logical synthesis, the value found for the standard version (without any of the optimizations done in this work) is 500MHz. The G-GPU has a similar performance across versions with different numbers of CUs because the CU itself is the bottleneck for performance in this architecture, not the logic for controlling the communication between the modules. As expected, the critical path for the version without any optimization has its starting point at a memory block. Also, the critical path was found inside the CU partition.
 
Larger memories, either in number of words or in word size, display a higher delay for accessing the stored data when compared with smaller memories. This observation guides our design space exploration: dividing the memory blocks in the critical path is a valid strategy for increasing the performance of a design~\cite{smartmem}. Memory division can be applied by diving the number of words, the size of the word, or both, the latter depending on the performance of the memories available in the given technology. This strategy requires a few alterations in the RTL code. First, the new modules have to be instantiated properly, substituting the target memories for the optimization. Second, the address or the input/output data have to be concatenated accordingly. To attain faster results, this task was fully automated in our framework. Thus, we only need to point which memories we want to divide and the number of divisions for applying this strategy.

Following our plan to achieve the best PPA ratio possible, we continually applied the memory division strategy when the critical path contained a memory block. However, the area of the memory blocks is not linear w.r.t. their size. In fact, two blocks of size $M\times N$ are larger and more power-hungry than a single block of size $2M\times N$ or $M\times 2N$. Therefore, from the division alone, we are increasing the area and power. Also, a small extra logic is necessary to accommodate the addressing control of the new blocks (i.e., MUXes to switch between block memories if the number of words is split according to the MSBs of the address). When exercising the memory division to enhance the design performance, we found cases where the critical path was not in memory blocks. For solving such timing issues, pipelines were introduced in those paths. In total, we created twelve different G-GPU solutions, varying the operating frequency and the number of CUs.
 
As a result, we created an open-source tool to automatically generate G-GPU IPs, from RTL to GDSII. The flow of GPUPlanner is highlighted in Fig.~\ref{fig:ip_flow}. For starters, the designer has to define the specifications required from the G-GPU. Our architecture can be configured for different numbers of CUs, ranging from 1 to 8. Increasing the number of CUs enhances the computation capacity of the G-GPU. Also, the designer has to specify the operating frequency of the G-GPU.

 After surveying the possible versions of the G-GPU for desired application scenarios, the designer can generate a specification for each scenario. Then, these specifications are contrasted with the characteristics of the technology intended to be used to create a first-order estimation of the G-GPU PPA. In this phase, there is a possibility to find several versions suitable for the given specification. Still, it also might happen that a configuration that suits the designer's requirements does not exist. However, our framework is not a static input generator. Instead, we provide a map on how to achieve a realistic PPA that might be close enough to the designer's requirements. This map is a dynamic spreadsheet, where the user input the delay of the memories blocks required for the non-optimized version of the G-GPU. Our map gives the maximum performance and which memory has to be divided or where to introduce pipelines to enhance the performance. This is an iterative process and can be repeated until the designer finds the desired performance. Thus, using our map, the designer can rapidly adapt his specification or create new versions of G-GPU by splitting more memory blocks to increase performance or by introducing on-demand pipelines. Even though applying this strategy is complicated, our framework can handle any memory and technology with little effort. The designer only has to give the basic information of the memory blocks (i.e., name, number of ports, port names, and minimum delay for data access). The only hard constraint is that many of the G-GPU memories have to be dual-port. Further development for single-port memories is scheduled as future work.

After settling the specifications, one or more designs can be feasible, generating a list of G-GPU versions. From a single push of a button, our framework can perform logic and physical synthesis of the list of designs. After the logic and physical synthesis, the resulting PPA is checked to guarantee it is under the initial specification. If the resulting G-GPU is out of the specifications, the designer should modify it and restart the process. In any case, the resulting layouts are ready to be integrated in a system as a tapeout-ready IP.

\begin{table*}[htb]
    \centering
    \caption{Characteristics of 12 different GGPU solutions generated by our tool after logic synthesis in Cadence Genus.}
    \begin{tabular}{p{1.8cm}p{2.2cm}p{2.6cm}p{1cm}p{1cm}p{1cm}p{1.8cm}p{1.8cm}p{1.4cm}}
    \hline
        \textbf{\#CU \& Freq. } &  \textbf{Total Area (mm$^2$)} & \textbf{Memory Area (mm$^2$)} & \textbf{\#FF} & \textbf{\#Comb.} & \textbf{\#Memory} & \textbf{Leakage (mW)} & \textbf{Dynamic (W)} & \textbf{Total (W)}  \\
        \hline
         1@500MHz & 4.19 & 2.68 & 119778 & 127826 & 51 & 4.62 & 1.97 & 2.055 \\
         2@500MHz & 7.45 & 4.64 & 229171 & 214243 & 93 & 8.54 & 3.63 & 3.77 \\
         4@500MHz & 13.84 & 8.56 & 437318 & 387246 & 177 & 16.07 & 6.88 & 7.14 \\ 
         8@500MHz & 26.51 & 16.39 & 852094 & 714256 & 345 & 30.79 & 13.33 & 13.86 \\ \hline
         1@590MHz & 4.66 & 3.15 & 120035 & 128894 & 68 & 4.73 & 2.57 & 2.66 \\
         2@590MHz & 8.16 & 5.34 & 229172 & 221946 & 120 & 8.73 & 4.63 & 4.81 \\
         4@590MHz & 15.03 & 9.72 & 436807 & 397995 & 224 & 16.41 & 8.70 & 9.02 \\
         8@590MHz & 28.65 & 18.49 & 850559 & 737232 & 432 & 31.25 & 16.81 & 17.40 \\ \hline
         1@667MHz & 4.77 & 3.26 & 120035 & 130802 & 71 & 4.65 & 2.62 & 2.72 \\
         2@667MHz & 8.27 & 5.45 & 229172 & 222028 & 123 & 8.72 & 4.69 & 4.87 \\
         4@667MHz & 15.15 & 9.83 & 436807 & 398124 & 227 & 16.43 & 8.75 & 9.07 \\
         8@667MHz & 28.69 & 18.60 & 848511 & 730506 & 435 & 30.21 & 19.10 & 19.76 \\
         \hline
    \end{tabular}

    \label{tab:ggpu_phys_char}
\end{table*}

\section{Results and Discussion}

During the exercise of the GPUPlanner in finding the best trade-offs for a range of operating frequencies, we were able to draw a map of parameters to be adapted to create the versions demonstrated in this work. This map is agnostic of the technology used because our main strategy of optimization deals with the intrinsic delay of the memories blocks and the characteristics of the G-GPU architecture. Employing our strategy for other technologies would result in different PPA ratios, depend on the given technology performance. The results depend mainly on the performance of the memories and of the standard cells. However, the points of optimization would be somewhat the same. If user follow our map, they will rapidly find the best versions for the given technology.
    
From the exercise of the GPUPlanner, we found 12 versions worth the PPA trade-off in a general manner. These versions have 1, 2, 4, and 8 CUs. Their variants run at 500MHz, 590MHz, and 667MHz. The characteristics of each version are shown in Table~\ref{tab:ggpu_phys_char}. In terms of area, the G-GPU size grows linearly with the number of CUs. The optimizations done for augmenting the performance increased the area by an average of 10\%, from 500MHz to 590MHz, and 2\%, from 590MHz to 667MHz. Thus, if the power consumption is not a priority, the 667MHz is a good fit for having a negligible increase in area in trade-off a better performance. These results demonstrate the potential scalability of the G-GPU architecture. 
    
After the logical synthesis, we chose four versions to perform the physical synthesis. Those are the 1CU@500MHz, 1CU@667MHz, 8CU@500MHz, and 8CU@677MHz. A reader can appreciate that these are the extreme cases identified by GPUPlanner. During this phase, the G-GPU is broken into three partitions during implementation: the CU, the general memory controller, and the top. The density of the CU and the general memory controller was set to 70\%. Because of our floorplan strategy of breaking the design into partitions, the top has a low density of 30\%. Nevertheless, breaking the design in partitions allows the designer to scale G-GPU without any extra effort. Once a CU partition is fully placed and routed, it can be implemented in versions with more than 1 CU by cloning the partition in the final floorplan of the design. Moreover, the user can create a collection of different CU layout blocks and scale the floorplan regarding the number of CUs for different application scenarios easily.
    
\begin{figure}[htb]
    \centering
    \includegraphics[width=0.875\columnwidth]{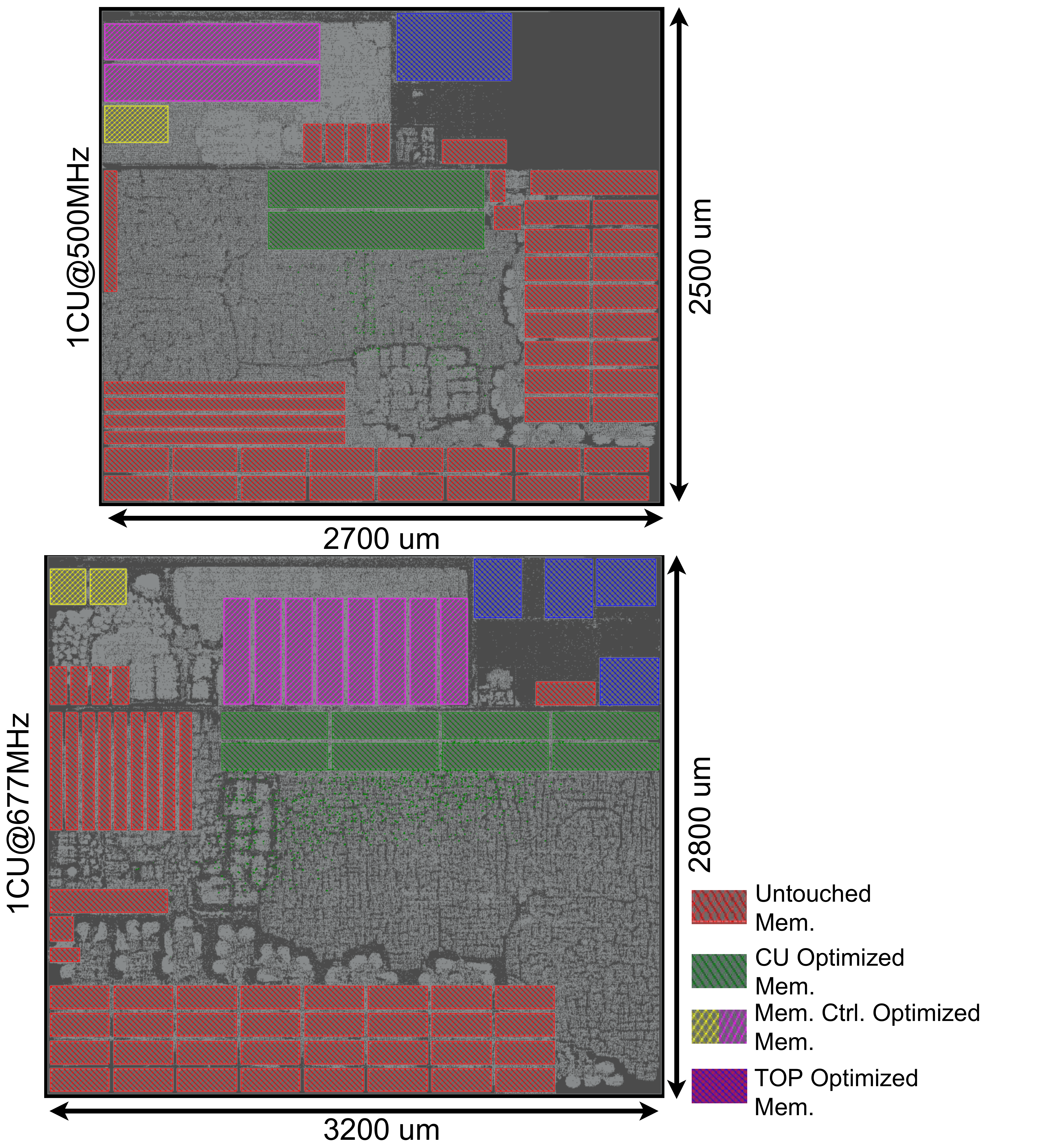}
    \caption{Layout comparison between 1CU@500MHz and 1CU@667MHz variants.}
    \label{fig:fgpu_layout_1cu}
\end{figure}

\begin{figure}[htb]
    \centering
    \includegraphics[width=0.90\columnwidth]{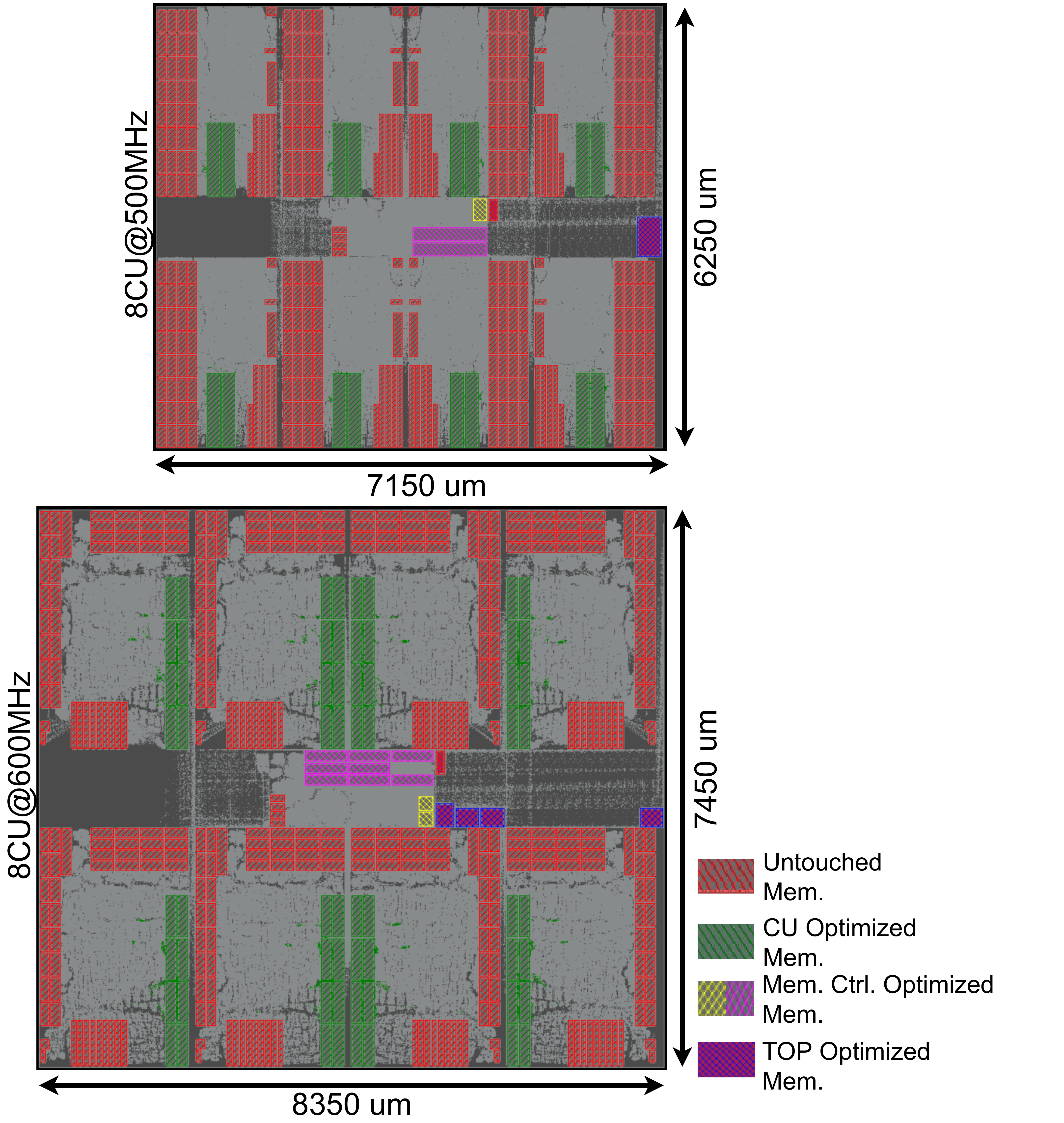}
    \caption{Layout comparison between minimum and maximum performance of a G-GPU with 8 CUs.}
    \label{fig:fgpu_layout_8cu}
\end{figure}

The layouts for the versions with 1 CU and 8 CUs are contrasted in  Fig.~\ref{fig:fgpu_layout_1cu} and Fig.~\ref{fig:fgpu_layout_8cu}, respectively. The block memories divided for augmenting the performance are highlighted in green for the CU partition, yellow and pink for the general memory controller, and blue for the top. Note how different the floorplan is between the version with optimizations running at 667MHz (600MHz in the 8 CUs version) and without optimizations running at 500MHz. Block memories have to be strategically placed in order to extract the maximum performance, hence, the differences in the floorplan. The layout of the versions 1CU@500MHz, 1CU@667MHz, 8CU@500MHz have the same performance expected from the logical synthesis (i.e., they can run at the specified clock frequency without any timing violation). However, the layout of version 8CU@667MHz can only run at 600MHz. This is explained by analyzing the floorplan of its layout (see Fig.~\ref{fig:fgpu_layout_8cu}).

The connecting routing wires introduce a significant capacitance because of the long distance between the peripheral CUs and the general memory controller. In turn, this capacitance increases path delay up to a point where it violates the 1.5ns target period.
To better explain the difference in wire length routing between 1 and 8 CUs, Table~\ref{tab:wl_cu_ggpu} shows the total amount of wire length per metal layer\footnote{For the technology utilized, the metal stack contains nine layers. The metal layers M1, M8, and M9 are reserved for power routing only and have not been drawn in Table~\ref{tab:wl_cu_ggpu}. This is a representative metal stack.}. In an attempt to solve this issue, pipelines were introduced between the connections with high delay, but this strategy was ineffective to solve the timing violations. For maintaining the PPA ratio balanced, the best performance we found for 8 CUs was 600MHz.

\begin{table}[htbp]
    \centering
    \caption{Routing length per metal layer  for different G-GPU versions and variants.}
    \begin{tabular}{p{0.3cm}p{1.5cm}p{1.5cm}p{1.5cm}p{1.6cm}}
        \hline
        \multicolumn{1}{c}{\textbf{Metal}} & \multicolumn{4}{c}{\textbf{Wirelength ($\mu m$)}} \\
        \multicolumn{1}{c}{\textbf{layer}} & \textbf{1CU@500MHz} & \textbf{1CU@667MHz} & \textbf{8CU@500MHz} & \textbf{8CU@600MHz} \\
        \hline
        M2 & 3185110  & 15340072 & 20314957 & 25637608   \\ 
        M3 & 5132356  & 21219705 & 27928578 & 34890963   \\
        M4 & 2987163  & 9866798  & 19209669 & 22387405  \\
        M5 & 2713788  & 11293663 & 21953276 & 26355211   \\
        M6 & 1430594  & 8801517  & 14074944 & 11111664   \\
        M7 &  616666  & 2915533  & 6316321  & 5315697    \\
        \hline 
    \end{tabular}
    \label{tab:wl_cu_ggpu}
\end{table}

  To fully evaluate the usage of G-GPU as an ASIC accelerator, we compared its performance with an implementation of the popular RISC-V architecture. We synthesized both architectures using the same technology utilized during the G-GPU implementation with an operating frequency of 667MHz, the RISC-V having 32kb memory and the G-GPU with its largest configuration for 1/2/4/8 CUs. As case-study applications, we chose seven micro-benchmarks from the AMD OpenCL SDK and increased their inputs up until crashing RISC-V and its compiler. We further increased the input size of the G-GPU applications to make its computing units fully utilized. To compare the performance of the different-input size applications, we took a pessimistic approach for G-GPU and considered that one could increase RISC-V application input sizes by multiplying its cycle count by the G-GPU/RISC-V input size ratio (which in practice is unfeasible but favors RISC-V). Table~\ref{tab:cycle_count} shows input sizes and measured cycle counts for all case-study applications.
  
\begin{figure}[ht]
    \centering
    \includegraphics[width=1\linewidth]{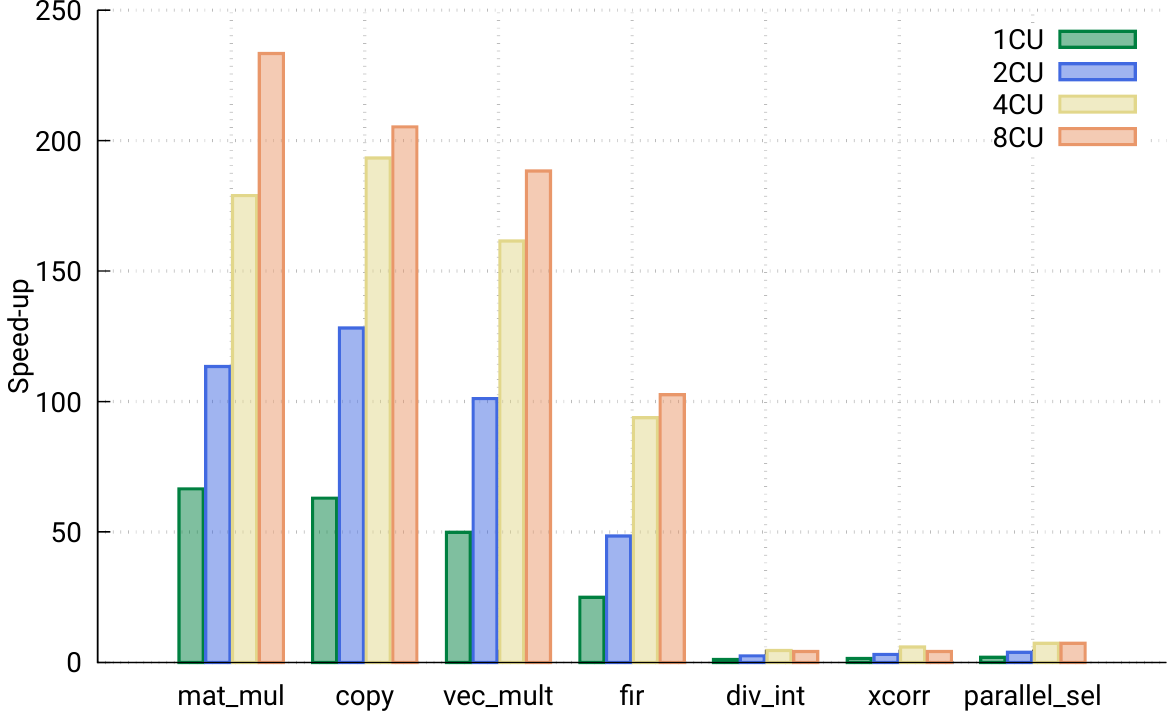}
    \caption{Speed-up over RISC-V.}
    \label{fig:raw_speedup}
\end{figure}

  Our first evaluation compares raw performance between G-GPU and RISC-V for the same input sizes. Fig.~\ref{fig:raw_speedup} shows in a bar chart that G-GPU with 8 CUs is up to 223 times faster than RISC-V. However, only applications that enjoy high parallelism are orders of magnitude faster when running using G-GPU. For applications with low to no parallelism, G-GPU can be as low as only 1.2 times faster than RISC-V. As G-GPU is a domain-specific ASIC accelerator, such results are expected, once it will not be the best option for general-purpose applications.
  Therefore, a user interested in implementing a G-GPU as an accelerator can utilize these provided data to ponder if this type of architecture is a good fit for his system when considering only the raw speed-up.
  
\begin{figure}[ht]
    \centering
    \includegraphics[width=1\linewidth]{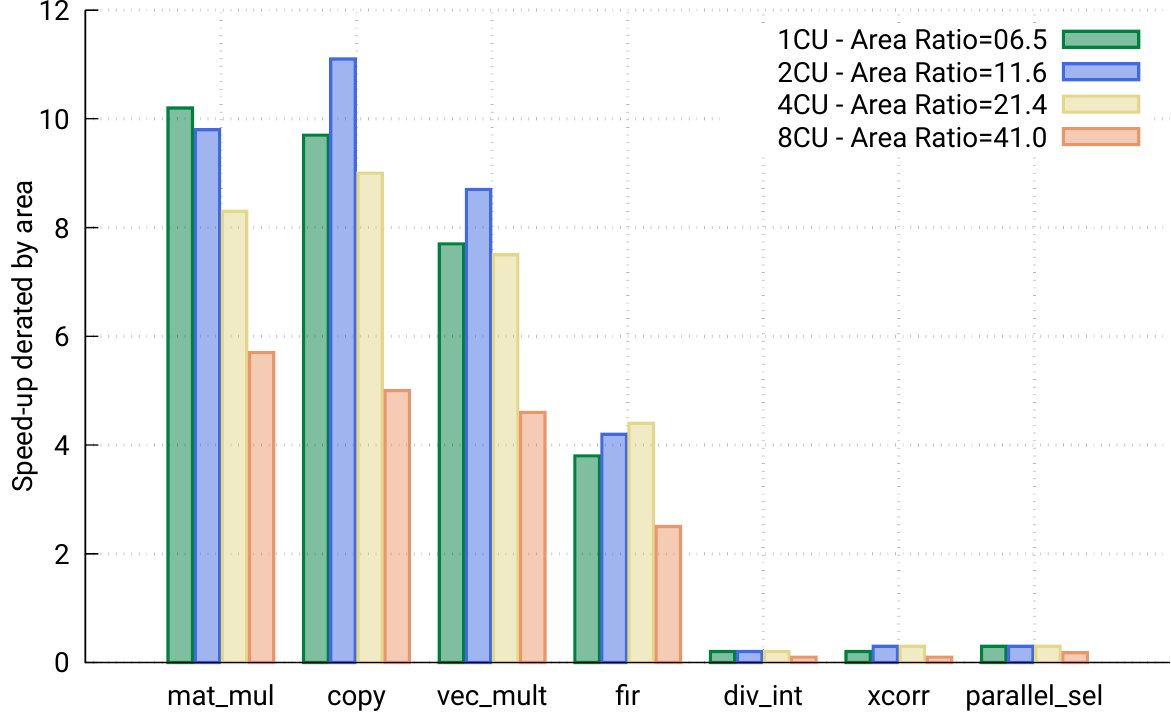}
    \caption{Speed-up over RISC-V derated by area.}
    \label{fig:area_speedup}
\end{figure}

  Our second evaluation factors previously measured area into performance speed-up. As designers might be interested in extracting the most out of a given available area, we derated the previously measured speed-up by dividing it by the G-GPU/RISC-V area ratio for each G-GPU CU configuration.
  This metric is useful to evaluate trade-offs in computation speed-up and area when replacing a RISV-C with a G-GPU. These results are shown in Fig.~\ref{fig:area_speedup} as a bar chart. G-GPU with 1 CU has an area that is 6.5 times larger than the RISC-V, and it achieves the best increase in performance per area of 10.2 times the RISC-V's. On the other hand, G-GPU with 8 CUs has an area that is 41 times bigger than RISC-V's, thus achieving the best increase in performance per area of 5.7 times faster than RISC-V's. Note that, when factoring area in, the 8-CU G-GPU shows the worst results. This trend happens mainly because data dependency and global memory communication limit parallelism. Thus, the provided increased processing power of a G-GPU configuration with more CUs.

%\begin{table}[h]
%\renewcommand{\arraystretch}{1.1}
%\caption{Case-study Applications}
%\label{tab:apps}
%\centering

%\begin{tabular}{p{1cm}p{1cm}p{1cm}p{1cm}p{1cm}p{1cm}p{1cm}p{1cm}p{1cm}}
%\hline
%\multirow{2}[4]{*}{Kernel} & \multicolumn{2}{p{2cm}}{Workload} &    & %\multicolumn{5}{p{3cm}}{Cycle Count (k) } \\
%\cmidrule{2-3}\cmidrule{5-9}   & RISC-V & FGPU &    & \multicolumn{1}{p{1cm}}{RISC-V} & \multicolumn{1}{p{1cm}}{1 CU} & \multicolumn{1}{p{1cm}}{2 CU} & \multicolumn{1}{c}{4 CU} & \multicolumn{1}{c}{8 CU} \\
%\midrule
%mat\_mul      & 128   & 2048  &     & 202 & 48   & 28   & 18   & 14 \\
%copy          & 512   & 32768 &     & 71  & 73   & 36   & 24   &  22 \\
%vec\_mul      & 1024  & 65536 &     & 78  & 100  & 49   & 31   & 26 \\
%fir           & 128   & 4096  &     & 542 & 694  & 358  & 185  & 169 \\
%div\_int      & 512   & 4096  &     & 32  & 209  & 105  & 57   & 62 \\
%xcorr         & 256   & 4096  &     & 542 & 5343 & 2802 & 1467 & 2079 \\
%paralle\_sel. & 128   & 2048  &     & 765 & 5979 & 3157 & 1656 & 1660 \\
%\hline
%\end{tabular}%

%\end{table}

\begin{table}[htb]
    \centering
    \caption{Benchmark's input size and cycle count}
    \begin{tabular}{p{0.9cm}p{0.7cm}p{0.8cm}p{0.8cm}p{0.5cm}p{0.5cm}p{0.5cm}p{0.5cm}}
    \hline
    \multirow{2}{*}{\textbf{Kerne}l} & \multicolumn{2}{c}{\textbf{Input size}} & \multicolumn{5}{c}{\textbf{Cycle Count (k-cycles})} \\
     & \textbf{RISC-V} & \textbf{G-GPU} & \textbf{RISC-V} & \textbf{1CU} & \textbf{2CU} & \textbf{4CU} & \textbf{8CU} \\
    \hline
mat\_mul      & 128   & 2048  & 202 & 48   & 28   & 18   & 14 \\
copy          & 512   & 32768 & 71  & 73   & 36   & 24   &  22 \\
vec\_mul      & 1024  & 65536 & 78  & 100  & 49   & 31   & 26 \\
fir           & 128   & 4096  & 542 & 694  & 358  & 185  & 169 \\
div\_int      & 512   & 4096  & 32  & 209  & 105  & 57   & 62 \\
xcorr         & 256   & 4096  & 542 & 5343 & 2802 & 1467 & 2079 \\
paralle\_sel  & 128   & 2048  & 765 & 5979 & 3157 & 1656 & 1660 \\
\hline
    \end{tabular}
    
    \label{tab:cycle_count}
\end{table}

For future work, we plan to update the GPUPlanner to be able to implement the 8-CU G-GPU with performance compared with the versions with fewer CUs. The performance problem of the layouts with 8 CUs has the possibility to be solved by replicating the general memory controller, shortening the distance between the peripheral CUs, and reducing the delay introduced by the routing wires. Also, we intend to scale FGPU beyond 8 CUs, including a supporting memory hierarchy, and incorporate single-port memories into GPUPlanner.

\section{Conclusion} \label{sec:conclusion}

In this work, we proposed a new solution for domain-specific ASIC accelerators based on GPU-like accelerators, called G-GPU. On top of that, we presented a framework -- GPUPlanner -- to fully automate the generation of G-GPUs from the RTL to a tape-out ready layout. Our results showed that G-GPUs are feasible domain-specific ASIC accelerator. Furthermore, when the G-GPU performance is contrasted with that of a RISC-V, it shows that our architecture has tremendous benefits for applications with high parallelism. Moreover, as GPUPlanner is an open-source framework, it gives the community the opportunity to explore the design space of GPU-like accelerators. Our work goes beyond the analysis of what constitutes a reasonable G-GPU accelerator in 65nm, as our tool can be easily extended to support other baseline GPU architectures and technologies. 

\bibliographystyle{ieeetr}
\bibliography{gpu}

\end{document}